\documentclass[twocolumn,prl,showpacs,floatfix,superscriptaddress]{revtex4-2}
\usepackage{graphicx}
\usepackage{microtype}
\usepackage{color}
\usepackage{amsmath}
\usepackage{stix}
\usepackage[T1]{fontenc}
\usepackage[colorlinks=true, linkcolor=blue, anchorcolor=black, citecolor=blue, filecolor=black, urlcolor=blue]{hyperref}
\usepackage[all]{hypcap}

\newcommand{\ikf}{Institut f\"ur Kernphysik, Goethe-Universit\"at, Max-von-Laue-Str. 1, D-60438 Frankfurt, Germany}
\newcommand{\ESRF}{ESRF, 6 Rue Jules Horowitz, BP 220, 38043 Grenoble Cedex 9, France}

\begin{document}
\title {
\large
Ion and Electron Momentum Distributions from Single and Double Ionization of Helium Induced by Compton Scattering
}
\author{M.~Kircher} \email{kircher@atom.uni-frankfurt.de} \affiliation{\ikf}
\author{F.~Trinter} \affiliation{\ikf} \address{Molecular Physics, Fritz-Haber-Institut der Max-Planck-Gesellschaft, Faradayweg 4-6, D-14195 Berlin, Germany}
\author{S.~Grundmann} \affiliation{\ikf}
\author{G.~Kastirke} \affiliation{\ikf}
\author{M.~Weller} \affiliation{\ikf}
\author{I.~Vela-Perez} \affiliation{\ikf}
\author{A.~Khan}  \altaffiliation{Institut f\"ur Ionenphysik und Angewandte Physik, Universit\"at Innsbruck, Technikerstraße 25/3, A-6020 Innsbruck, Austria}\affiliation{\ikf}
\author{C.~Janke} \affiliation{\ikf}
\author{M.~Waitz} \affiliation{\ikf}
\author{S.~Zeller} \affiliation{\ikf}
\author{T.~Mletzko} \affiliation{\ikf} 
\author{D.~Kirchner} \affiliation{\ikf}
\author{V.~Honkim\"aki}\affiliation{\ESRF}
\author{S.~Houamer} \affiliation{LPQSD, Department of Physics, Faculty of Science, University S\'etif-1, 19000, Setif, Algeria}
\author{O.~Chuluunbaatar} \affiliation{Joint Institute for Nuclear Research, Dubna, Moscow region 141980, Russia}
\address{Institute of Mathematics and Digital Technology, Mongolian Academy of Sciences, 13330, Ulaanbaatar, Mongolia}
\author{Yu.~V.~Popov} \affiliation{Skobeltsyn Institute of Nuclear Physics, Lomonosov Moscow State University, Moscow 119991, Russia}\affiliation{Joint Institute for Nuclear Research, Dubna, Moscow region 141980, Russia}
\author{I.~P.~Volobuev} \affiliation{Skobeltsyn Institute of Nuclear Physics, Lomonosov Moscow State University, Moscow 119991, Russia}
\author{M.~S.~Sch\"offler} \affiliation{\ikf}
\author{L.~Ph.~H.~Schmidt}\affiliation{\ikf}
\author{T.~Jahnke} \affiliation{\ikf}
\author{R.~D\"orner} \email{doerner@atom.uni-frankfurt.de}
\address{\ikf}

\begin{abstract}
\noindent
We present the momentum distributions of the nucleus and of the electrons from double ionization of the helium atom by Compton scattering of photons with $h\nu=40$\,keV. We find that the doubly charged ion momentum distribution is very close to the Compton profile of the nucleus in the ground state of the helium atom, and the momentum distribution of the singly charged ion to give a precise image of the electron Compton profile. To reproduce these results, non-relativistic calculations require the use of highly correlated initial- and final-state wavefunctions.
\end{abstract}
\maketitle

Compton scattering is one of the fundamental interaction processes of light with matter. In his seminal paper reporting the discovery of the process, A. Compton described it as a binary collision between a photon and one single quasi-free electron at rest. Soon after, initial momenta of the kicked electron where included into the description \cite{DuMond1929} leading to what is known today as the impulse approximation. This has established Compton scattering as a tool to study momentum distributions of single active electrons in bound states, often referred to as Compton profiles (see \cite{Cooper1985} for a review). In this quasi-free electron approximation, the kinematics are easily derived from momentum and energy conservation and the cross section is given by the Klein-Nishina equation \cite{KleinNishina1929}. If the binding energy of the electron plays a role, Compton scattering becomes more intriguing. Comprehensive experiments for such more general conditions are scarce even today due to the very small cross section. Coincidence experiments on single electron processes have been reported recently in Ref.\,\cite{Kircher2020}. 

The next frontier for studies of Compton scattering are situations where the single-active-electron approximation fails and more electrons are actively involved. These events are mediated by electron-electron interaction and have therefore been discussed as an experimental approach to explore correlated bound states. Such two-electron Compton events are usually dissected in two steps: the Compton event occurring at only one electron which is kicked by the photon and the second electron is ionized by electron-electron interaction. Depending on the order of these interactions, the process is referred to as knock-off or shake-off \cite{McGuire1997,Samson1990,Knapp2002a}. For knock-off, the Compton event occurs first and the Compton electron is pictured as to ``kick'' the other electron on its way out, transferring part of its energy and momentum. For shake-off, the electron-electron interaction is part of the ground state and the sudden removal of one electron leads to a shake-off of the second one to the continuum. These correlation-driven processes also occur for photoabsorption, electron, or ion impact (see, e.g., \cite{McGuire1995} and references therein). The probabilities and even more so the differential cross sections for these processes, however, are very different in all cases. The main reasons for the differences between the ionization schemes are angular-momentum and parity selection rules, and the fact that different parts of the bound-state wavefunction are affected by different types of interactions. Angular-momentum selection rules, which are strict for single-photon absorption, strongly shape the angular distribution \cite{Maulbetsch1995} masking fingerprints of electron correlation in the final-state momenta of the emitted electrons. Furthermore, single-photon absorption at high energies is selective to high-momentum components in the initial state, thus probes only a very specific component of the wavefunction. In these respects, Compton scattering clearly stands out, as at high energies, the Compton scattering probability does not depend on the initial position or momentum of the electron while it is bound. There are also no angular-momentum or parity selection rules for Compton scattering. Despite this fundamental nature, however, ejection of two electrons by Compton scattering has been studied experimentally only on the level of total cross sections \cite{Spielberger1996,Krassig1999}. There, it has been established that the ratio of double to single ionization of helium by Compton scattering approaches $R^\infty_c=0.86\%$ for high photon energies, as compared to $R_\gamma^\infty=1.66\%$ for photoabsorption \cite{Andersson1993} and $R_{\textit{charged}}^\infty=0.26\%$ for charged particle impact \cite{McGuire1995}. This asymptotic value is, however, approached very slowly with increasing energy for Compton scattering \cite{Spielberger1999}. 

In the present paper, we present the first differential study of two-electron Compton scattering and compare its momentum balance with the case of single ionization. We choose helium double ionization as the cleanest possible process of this kind. We find for single ionization that the left-behind ion shows a very clear fingerprint of the single-electron bound-state momentum profile including its modification by initial-state correlation, and, surprisingly, for Compton-scattering-driven double ionization, the continuum momentum distribution of the ion mimics the momentum distribution of the nucleus in the helium bound state.

In our experiment, we induce helium single or double ionization with an x-ray photon of an energy $h\nu$ = 40\,keV:
\begin{align}
 h\nu + \mathrm{He} &\rightarrow h\nu' + \mathrm{He^{1+}} + e^- \ , \label{eq1}\\
 h\nu + \mathrm{He} &\rightarrow h\nu' + \mathrm{He^{2+}} + e^-_1 + e^-_2 \ . \label{eq3}
\end{align}
The momenta of the respective particles of the reactions (\ref{eq1}) and (\ref{eq3}) are given by
\begin{align}
 \vec k_{\gamma} + \vec p_\mathrm{He} &=  \vec k_{\gamma'} + \vec p_\mathrm{He^{1+}} + \vec p_{e} \ ,  \label{eq2} \\
 \vec k_{\gamma} + \vec p_\mathrm{He}  &=  \vec k_{\gamma'} + \vec p_\mathrm{He^{2+}} + \vec p_{e1} +\vec p_{e2} \ , \label{eq4}
\end{align}
respectively, and are related by momentum conservation. Single and double ionization are induced by the momentum transfer $\vec{Q} = \vec{k}_{\gamma} - \vec{k}_{\gamma'}$. In our experiment, we detect the ion charge state and the 3D momentum $\vec p_\mathrm{He^{q+}}$ of the ion and, in case of double ionization, the 3D momentum $\vec p_{e2}$ of one of the two electrons. We detect only electrons with a momentum magnitude $p_{e2}<1.1$\,a.u. Since for single ionization the majority of electrons have momentum magnitudes larger than 1.1\,a.u., we do not detect the electron for single ionization. For double ionization, we only detect slow electrons. We do not detect the scattered photon.

The cross section for ionization of helium by Compton scattering at 40\,keV photon energy is only on the order of $10^{-24}$\,$\mathrm{cm}^2$ \cite{Samson1994}. The cross section for double ionization is less than 1\% of this. Therefore, performing such an experiment in the gas phase requires the combination of highly efficient detection methods and high-intensity light sources. The experiment reported here was performed at beam line ID31 of the European Synchrotron Radiation Facility (ESRF) in Grenoble, France, using a cold target recoil ion momentum spectroscopy (COTLRIMS) reaction microscope \cite{Ullrich2003}. A supersonic helium gas jet was crossed with linearly polarized synchrotron light at right angle within a COLTRIMS spectrometer. A pinhole monochromator \cite{Vaughan2011} was used to select the photon energy of $h\nu=40$\,keV. An argon filter unit after the undulator removes low-energy photons from the beam. The overlap between gas jet and photon beam defines a localized reaction region of approximately $0.4 \times 0.1 \times 1.0\,\mathrm{mm^3}$. The synchrotron machine operated with 16 electron bunches in the storage ring at 5.68\,MHz bunch rate, with a photon flux of $8.4\times 10^{14}$\,photons/s at $\Delta E/ E=1.1\%$. Electric (6.5\,V/cm) and magnetic (6.4\,Gs) fields within the spectrometer guide the charged reaction fragments onto two position- and time-sensitive microchannel plate detectors with a delay-line anodes \cite{Jagutzki2002}. The electron side of the spectrometer had a total length of 31.2\,cm, divided in an acceleration and a drift region with a length ratio of 1:2 (time-of-flight focusing). The ion side had a total length of 146\,cm and included an electrostatic lens to compensate for the finite size of the reaction region \cite{Dorner1997}. We utilized $\mathrm{He^{2+}}$ ion beams of 20 to 50\,keV energy to calibrate our experiment. An electron capture process $\mathrm{He^{2+}}+\mathrm{He}\rightarrow \mathrm{He^{1+}}+\mathrm{He^{1+}}$ was used to calibrate our ion detector. Electrons with well defined energies produced by the autoionization channel of the reaction $\mathrm{He^{2+}} + \mathrm{Ne} \rightarrow \mathrm{He^{2+}} + \mathrm{Ne^{1+}} + e^-$ have been measured to calibrate the electron detector and the electric and magnetic fields of the spectrometer (see \cite{Ullrich2003} for the kinematics of ion collisions). The uncertainty in the $\mathrm{He^{2+}}$ projectile velocity yields a systematic uncertainty of our measured momenta of about 1.5\%.

We compare our results for double ionization with non-relativistic calculations using the $A^2$ approximation where the transition matrix element $M=M(\vec p_{e1}, \vec p_{e2}; \vec Q)$ is given by
\begin{align}
M(\vec p_{e1}, \vec p_{e2}; \vec Q) = (\vec \epsilon \cdot \vec \epsilon')\langle \Phi_\textit{f}\ |e^{i\vec Q \cdot \vec r_1}+e^{i\vec Q\cdot \vec r_2}|\Phi_i\rangle \ .   \label{eq5}
\end{align}
$\vec r_1$ and $\vec r_2$ are the positions of the  electrons, and $\vec\epsilon$ and $\vec\epsilon'$ are the polarization vectors of the incoming and outgoing photons, respectively.  $\Phi_\textit{f,i}$ denotes the final- and initial-state wavefunction. 

The $A^2$ approximation for single ionization describes ionization by Compton scattering as the product of scattering of the photon at a free electron, described by the Klein-Nishina cross section leading to momentum transfer $\vec Q$, times the overlap of the bound-state wavefunction shifted in momentum space by $Q$, with the continuum. Analogously, Eq.~(\ref{eq5}) describes double ionization in which electron~1 and electron~2 are kicked coherently with momentum transfer $Q$, and the overlap of the momentum-shifted two-electron ground state with the two-electron continuum is evaluated.

\begin{figure}[b]
\includegraphics{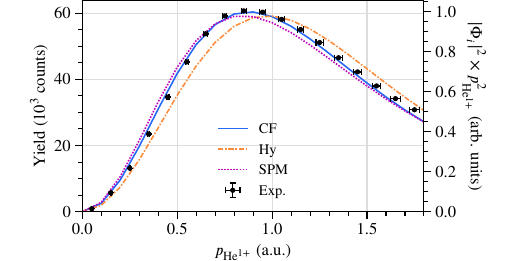}
\caption{Momentum distribution of $\mathrm{He^{1+}}$ ions produced by Compton scattering of photons with $h\nu= 40$\,keV. The experimental yield is normalized to the integral of the solid blue line. The horizontal error bars of the data points give the systematic accuracy of our momentum calibration. The vertical error bars are the standard statistical error. The lines give the bound-state electron momentum distribution (Compton profile) for different helium ground-state wavefunctions, namely Hylleraas (Hy), correlated trial (CF), and configuration-interaction (SPM) wavefunctions. See text for a discussion thereof.}
\label{fig1}
\end{figure}
We first discuss the result for the well understood case of single ionization. Figure\,\ref{fig1} shows the momentum distribution of the $\mathrm{He^{1+}}$ ions. The lines show the predicted bound-state momentum distribution of one electron for three different helium ground-state wavefunctions. The solid blue line corresponds to a highly correlated trial wavefunction (CF) \cite{Chuluunbaatar2006} for which the calculated binding energy equals the true value to 7 significant digits, the dash-dotted orange line corresponds to a single-parameter Hylleraas wavefunction (Hy) \cite{Hylleraas1928} for Z=27/16, resulting in a binding energy of 2.85\,a.u.\ (as compared to the true value of 2.903724\,a.u.), and the dotted magenta line to the most simple configuration-interaction wavefunction (SPM, first discussed by Silverman, Platas, and Matsen \cite{Silverman1960}) which includes only a small amount of $(2p)^2$ with a binding energy of 2.8952278 a.u. The experimental $\mathrm{He^{1+}}$ ion momentum distribution is within our calibration accuracy in excellent agreement with the electron momentum distribution from the best wavefunction (that is, CF, the wavefunction that includes the highest degree of correlations). We note that the horizontal error bars in the histogram do not reflect statistical errors, but give the systematical uncertainly of our momentum calibration of 1.5\%, thus show the effect of an overall stretch or compression of the horizontal axis. The close match of the ion momentum distribution with the bound-electron Compton profile confirms the validity of the impulse approximation (compare \cite{Kaliman2004} for theory). Compton scattering transfers the momentum $\vec Q$ to the electron. If the corresponding energy $Q^2/2m_e$ is large compared to the binding energy, the ionization probability becomes independent of the bound momentum of the electron (compare \cite{Kircher2020} for the other extreme) and the Coulomb potential can be neglected for the kinematics. In this case, as described by the impulse approximation, for a single active electron the ion momentum distribution is the exact mirror image of the bound-electron momentum distribution. The present experimental conditions come close to this ideal situation. According to the Klein-Nishina cross section, at 40\,keV photon energy, only 0.7\% of all Compton events correspond to an energy transfer below the helium binding energy of 24.6\,eV. Furthermore, two-electron effects play only a minor role for the data in Fig.\,\ref{fig1}, since the $\mathrm{He^{1+}}$ momentum distribution contains all ground- and excited-state ions and only a very small fraction of about 1\% of the Compton events are not included as they lead to double ionization.

\begin{figure}
\includegraphics{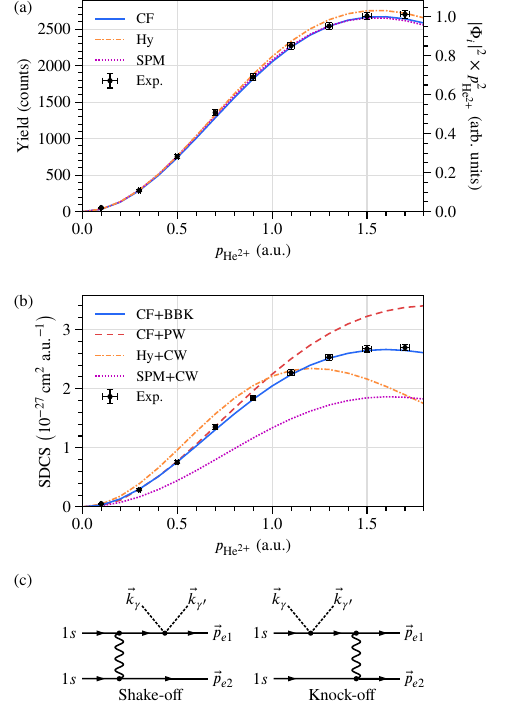}
\caption{Momentum distribution of $\mathrm{He^{2+}}$ ions produced by Compton scattering of photons at $h\nu=40$\,keV. Panels (a) and (b) show the same experimental data points. The data points are normalized to the integral of the solid blue line, respectively. The horizontal error bars correspond to the systematic accuracy of our momentum calibration. The vertical error bars are the standard statistical error. (a) Comparison of the nuclear Compton profile (the momentum distribution of the nucleus in bound helium). The lines show the different initial ground-state wavefunctions. (b) Comparison of our data with the calculated singly differential cross section (SDCS) for different sets of initial ground-state (CF, Hy, SPM) and final-state (BBK, CW, PW) wavefunctions (see text for a comprehensive explanation of all lines). (c) Feynman diagrams describing double ionization by Compton scattering (adapted from \cite{Amusia1995}). In (c), two Feynman diagrams with permutations of lines $\vec p_{e1}$ and $\vec p_{e2}$ were omitted.}
\label{fig2}
\end{figure}
In the remainder of this letter, we discuss the most interesting channel of double ionization. The corresponding momentum distribution of the $\mathrm{He^{2+}}$ nucleus (Fig.\,\hyperref[fig2]{2(a)}) is much broader than the $\mathrm{He^{1+}}$ momentum distribution. As Fig.\,\hyperref[fig2]{2(a)} shows, the measured distribution matches the momentum distribution (Compton profile) of the nucleus (solid blue line) in the bound helium atom prior to the photon impact. This indicates the validity of a kind of sudden approximation for the electron-pair removal. For this, two ingredients are necessary: (i) two-electron Compton scattering needs to remove both electrons from the initial state without altering the nucleus momentum, as it is assumed in the sudden approximation for one electron; and (ii) the two-electron removal must provide an unbiased sampling of the full wavefunction without selecting specific regions in momentum space. Ingredient (i) suggests that in the Feynman diagrams shown in Fig.\,\hyperref[fig2]{2(c)}, the photon-electron vertex, as well as the electron-electron vertex, describe interactions to which the nucleus is a spectator only and is not affected in its momentum. Note that the inclusion of the Feynman diagrams is for visualization of the process only, they are not the basis of our theoretical calculations. The passive role of the nucleus in the electron-electron interaction is also underlined by the similarity between nuclear momentum distributions from the different quality bound states shown by the lines in Fig.\,\hyperref[fig2]{2(a)}. Despite the significantly different amount of electron-electron correlations, which lead to the differences in the single-electron momentum distributions as shown in Fig.\,\ref{fig1}, the nuclear Compton profile in Fig.\,\hyperref[fig2]{2(a)} is very similar for all the wavefunctions, showing that the electron-electron interaction is quite decoupled from the nucleus. Ingredient (ii)---the unbiased sampling of the initial state---is a finding which is even more surprising in the light of previous literature. For example, it is known that the shake-off probability depends on the momentum of the electron which is removed from the two-electron ground state \cite{Shi2002}. This gives rise to the different high-energy asymptotes of ratios of double to single ionization for Compton scattering and photoabsorption \cite{Aberg1970,Spielberger1995} and their relation to charged-particle impact. 

Figure\,\hyperref[fig2]{2(b)} compares the measured momentum distributions with our non relativistic calculation. We have evaluated the matrix element in Eq.\,(\ref{eq5}) with four sets of initial and final states. For the initial state, we have used wavefunctions with different degrees of correlation, as shown in Figs.\,\ref{fig1} and \hyperref[fig2]{2(a)}. For the final state, we have used the fully correlated Brauner-Briggs-Klar (BBK) wavefunction \cite{Brauner1989} which accounts for the electron-nucleus and the electron-electron interaction on equal footing \cite{Briggs2016}. For calculations of double ionization by photoabsorption \cite{Briggs2000} and for ($e$,2$e$) processes \cite{Brauner1989}, this fully correlated final-state wavefunction has been shown to yield extremely good results also for very correlation-sensitive observables, such as the distribution of the mutual angle between the electrons. Our ``best'' calculation using the highly correlated initial state and the BBK wavefunction for the finals state (solid blue line in Fig.\,\hyperref[fig2]{2(b)}) yields extremely good agreement with our experimental data. All the other calculations (labeled CF+PW, Hy+CW, SPM+CW in the figure legend), where we have used final states which do not account for electron-electron repulsion---namely, a Coulomb wave with $Z=2$ for electron~2 and either a Coulomb wave with $Z=1$ (CW) or a plane wave (PW) for electron~1---are significantly off. In terms of the Feynman diagrams (Fig.\,\hyperref[fig2]{2(c)}), using final states without an electron-electron interaction term corresponds to neglecting the knock-off process. This underlines that at the present photon energy the shake-off limit is not yet reached. 

\begin{figure}
\includegraphics{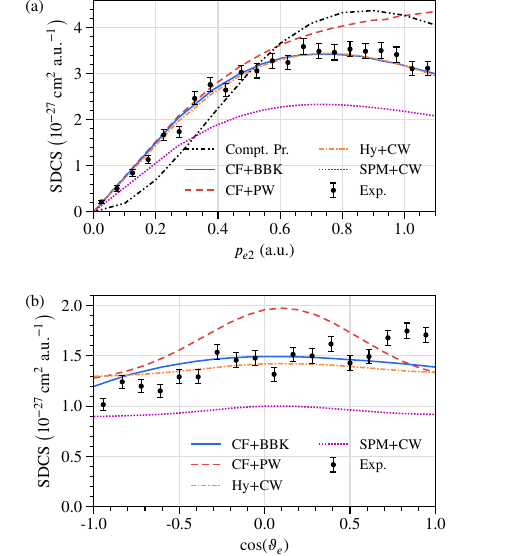}
\caption{Double ionization by Compton scattering of photons at $h\nu=40$\,keV. Respectively, the single differential cross sections for electron momenta (a) and the emission angle $\vartheta_e$ between the electron and incident photon beam direction (b) are shown. The data are integrated over 0.1--1.1\,a.u.\ electron momentum and normalized to the integral of the blue solid line, respectively. The error bars are the standard statistical error. Colored lines are calculations using different initial and final states (same as in Fig.\,\hyperref[fig2]{2(b)}). The dash-dot-dotted black line is the momentum distribution of the bound electron.}
\label{fig3}
\end{figure}
To further elucidate the role of electron correlations on the observables, we now inspect the momentum and angular distribution of the emitted electrons (Fig.\,\ref{fig3}). Comparison with the electron momentum distribution in the ground state (see Fig.\,\ref{fig1}, shown in Fig.\,\hyperref[fig3]{3(a)} as dash-dot-dotted black line), show that the continuum electrons have significantly smaller momenta. This is expected as, e.g., for the shake-off as well as for the knock-off mechanism a small energy transfer to the secondary electron is favored. The experimental electron-momentum distribution (Fig.\,\hyperref[fig3]{3(a)}) is very well described by our calculation using the fully correlated initial and final state. When we remove the knock-off process by using a plane wave for the final state while keeping the fully correlated initial state (dashed red line labeled CF+PW) we find large discrepancies at higher electron momenta, which is in agreement with the findings in Ref.\,\cite{Knapp2002a}. The good agreement of the calculations using the Hylleraas initial wavefunction and two final Coulomb wavefunctions for the electrons we suspect to be incidental. In Fig.\,\hyperref[fig3]{3(b)}, we find the electron angular distribution to be almost isotropic. We note that this angular distribution is integrated over all electron momenta from 0.1 to 1.1\,a.u., where the lower bound is due to our experimental momentum resolution. For photon scattering angles below 15\,deg, corresponding to 3\% of the Klein-Nishina cross section, the energy transfer $Q^2/(2m_e)$ is below 111\,eV (which is the sum of the double-ionization threshold of 79\,eV and continuum energies of $16\mathrm{\,eV}=(1.1\mathrm{\,a.u.})^2/(2m_e)$ per electron). For those cases, the Compton scattered electron and the secondary electron are indistinguishable and both contribute to the events in Figs.\,\hyperref[fig3]{3(a)} and \hyperref[fig3]{3(b)}. In these cases, $\vec Q$ is almost perpendicular to the photon propagation direction. In the calculation using a plane wave for the faster electron, these low-momentum-transfer double-ionization events therefor lead to the peak visible at $\cos(\vartheta) \simeq 0.1$. In reality, the Coulomb interaction with the nucleus and, even more, the electron-electron interaction redistributes these electrons in angle. This is in accordance with, e.g., the findings from ($\gamma$,2$e$) experiments where, at low energies, the electrons are not directed strongly along the polarization axis, and the electrons show a $\beta$ parameter around zero \cite{Wehlitz1991,Knapp2002b}.

In conclusion, we have presented the differential measurement of a one-electron and a two-electron Compton scattering process. We find that with single ionization the ion momentum distributions are a mirror image of the bound electron Compton profile which is, as the comparison with our calculations shows, highly sensitive to electron-electron correlations in the bound state. Two-electron Compton scattering is found to project the unperturbed bound-state momenta of the nucleus to the continuum, making them directly observable. 

\begin{acknowledgments}
This work is supported by DFG (Deutsche Forschungsgemeinschaft). We thank H. Isern, J. Drnec, and F. Russello from beam line ID31 at ESRF for excellent support during the beam time. The calculations were performed on the basis of the heterogeneous computing platform HybriLIT on supercomputer ``Govorun'' (LIT, JINR). The work was partially supported by the Hulubei-Meshcheryakov JINR programs, grant of RFBR and MECSS No. 20-51-4400, grant of Foundation of Science and Technology of Mongolia SST 18/2018, grant of the RFBR No. 19-02-00014a. S.H. thanks the DGRSDT-Algeria Foundation for support. A.K. acknowledges support by the Alexander von Humboldt Foundation.
\end{acknowledgments}


\begin{thebibliography}{99}
\bibitem{DuMond1929} J. W. M. Du\,Mond, \href{https://doi.org/10.1103/PhysRev.33.643}{Phys. Rev. \textbf{33}, 643 (1929).}
%
\bibitem{Cooper1985} M. J. Cooper, \href{https://doi.org/10.1088/0034-4885/48/4/001}{Rep. Prog. Phys. \textbf{48}, 415 (1985).}
%
\bibitem{KleinNishina1929} O. Klein and Y. Nishina, \href{https://doi.org/10.1007/BF01366453}{Z. Phys. \textbf{52}, 853--868 (1929).}
%
\bibitem{Kircher2020} M. Kircher, F. Trinter, S. Grundmann, I. Vela-Perez, S. Brennecke, N. Eicke, J. Rist, S. Eckart, S. Houamer, O. Chuluunbaatar \textit{et al.}, \href{https://doi.org/10.1038/s41567-020-0880-2}{Nat. Phys. \textbf{16}, 756--760 (2020).}
%
%
\bibitem{McGuire1997} J. H. McGuire, \textit{Electron Correlation Dynamics in Atomic Collisions} (Cambridge University Press, Cambridge, UK, 1997).
%
\bibitem{Samson1990} J. A. R. Samson, \href{https://doi.org/10.1103/PhysRevLett.65.2861}{Phys. Rev. Lett. \textbf{65}, 2861 (1990).}
%
\bibitem{Knapp2002a} A. Knapp, A. Kheifets, I. Bray, Th. Weber, A. L. Landers, S. Sch\"ossler, T. Jahnke, J. Nickles, S. Kammer, O. Jagutzki \textit{et al.}, \href{https://doi.org/10.1103/PhysRevLett.89.033004}{Phys. Rev. Lett. \textbf{89}, 033004 (2002).}
%
\bibitem{McGuire1995} J. H. McGuire, N. Berrah, R. Bartlett, J. A. R. Samson, J. A. Tanis, C. L. Cocke, and A. S. Schlachter, \href{https://doi.org/10.1088/0953-4075/28/6/009}{J. Phys. B: At. Mol. Opt. Phys. \textbf{28}, 913 (1995).}
%
\bibitem{Maulbetsch1995} F. Maulbetsch and J. S. Briggs, \href{https://doi.org/10.1088/0953-4075/28/4/007}{J. Phys. B: At. Mol. Opt. Phys. \textbf{28}, 551 (1995).}
%
\bibitem{Spielberger1996} L. Spielberger, O. Jagutzki, B. Kr\"assig, U. Meyer, Kh. Khayyat, V. Mergel, Th. Tschentscher, Th. Buslaps, H. Br\"auning, R. D\"orner \textit{et al.}, \href{https://doi.org/10.1103/PhysRevLett.76.4685}{Phys. Rev. Lett. \textbf{76} 4685 (1996).}
%
\bibitem{Krassig1999} B. Kr\"assig, R. W. Dunford, D. S. Gemmell, S. Hasegawa, E. P. Kanter, H. Schmidt-B\"ocking, W. Schmitt, S. H. Southworth, Th. Weber, and L. Young, \href{https://doi.org/10.1103/PhysRevLett.83.53}{Phys. Rev. Lett. \textbf{83}, 53 (1999).}
%
\bibitem{Andersson1993} L. R. Andersson and J. Burgd\"orfer, \href{https://doi.org/10.1103/PhysRevLett.71.50}{Phys. Rev. Lett. \textbf{71}, 50 (1993).}
%
\bibitem{Spielberger1999} L. Spielberger, H. Br\"auning, A. Muthig, J. Z. Tang, J. Wang, Y. Qiu, R. D\"orner, O. Jagutzki, Th. Tschentscher, V. Honkim\"aki \textit{et al.}, \href{https://doi.org/10.1103/PhysRevA.59.371}{Phys. Rev. A \textbf{59}, 371 (1999).}
%
\bibitem{Samson1994} J. A. R. Samson, Z. X. He, R. J. Bartlett, and M. Sagurton, \href{https://doi.org/10.1103/PhysRevLett.72.3329}{Phys. Rev. Lett. \textbf{72}, 3329 (1994).}
%
\bibitem{Ullrich2003} J. Ullrich, R. Moshammer, A. Dorn, R. D\"orner, L. Ph. H. Schmidt, and H. Schmidt-B\"ocking, \href{https://doi.org/10.1088/0034-4885/66/9/203}{Rep. Prog. Phys. \textbf{66} 1463 (2003).}
%
\bibitem{Vaughan2011} G. B. M. Vaughan, J. P. Wright, A. Bytchkov, M. Rossat, H. Gleyzolle, I. Snigireva, and A. Snigirev, \href{https://doi.org/10.1107/S0909049510044365}{J. Synchrotron Radiat. \textbf{18}, 125 (2011).}
%
\bibitem{Jagutzki2002} O. Jagutzki, A. Cerezo, A. Czasch, R. D\"orner, M. Hattas, Min Huang, V. Mergel, U. Spillmann, K. Ullmann-Pfleger, T. Weber \textit{et al.}, \href{https://doi.org/10.1109/TNS.2002.803889}{IEEE Trans. Nucl. Sci. \textbf{49}, 2477 (2002).}
%
\bibitem{Dorner1997} R. D\"orner, V. Mergel, L. Spielberger, M. Achler, Kh. Khayyat, T. Vogt, H. Br\"auning, O. Jagutzki, T. Weber, J. Ullrich \textit{et al.}, \href{https://doi.org/10.1016/S0168-583X(96)00877-4}{Nucl. Instr. and Meth. in Phys. Res. B \textbf{124}, 225--231 (1997).}
%
\bibitem{Chuluunbaatar2006} O. Chuluunbaatar, I. V. Puzynin, P. S. Vinitsky, Yu. V. Popov, K. A. Kouzakov, and C. Dal Cappello, \href{https://doi.org/10.1103/PhysRevA.74.014703}{Phys. Rev. A \textbf{74}, 014703 (2006).}
%
\bibitem{Hylleraas1928} E. A. Hylleraas, \href{https://doi.org/10.1007/BF01340013}{Z. Phys. \textbf{48}, 469--494 (1928).}
%
\bibitem{Silverman1960} J. N. Silverman, O. Platas, and F. A. Matsen, \href{https://doi.org/10.1063/1.1730930}{J. Chem. Phys. \textbf{32}, 1402 (1960).}
%
\bibitem{Kaliman2004} Z. Kaliman and K. Pisk, \href{https://doi.org/10.1016/j.radphyschem.2004.04.035}{Rad. Phys. Chem. \textbf{71}, 633--635 (2004).}
%
\bibitem{Amusia1995} M. Ya. Amusia and A. I. Mikhailov, \href{https://doi.org/10.1088/0953-4075/28/9/011}{J. Phys. B: At. Mol. Opt. Phys. \textbf{28}, 1723 (1995)}
%
\bibitem{Shi2002} T. Y. Shi and C. D. Lin, \href{https://doi.org/10.1103/PhysRevLett.89.163202}{Phys. Rev. Lett. \textbf{89}, 163202 (2002).}
%
\bibitem{Aberg1970} T. \r{A}berg, \href{https://doi.org/10.1103/PhysRevA.2.1726}{Phys. Rev. A \textbf{2}, 1726 (1970).}
%
\bibitem{Spielberger1995} L. Spielberger, O. Jagutzki, R. D\"orner, J. Ullrich, U. Meyer, V. Mergel, M. Unverzagt, M. Damrau, T. Vogt, I. Ali \textit{et al.}, \href{https://doi.org/10.1103/PhysRevLett.74.4615}{Phys. Rev. Lett. \textbf{74}, 4615 (1995).}
%
\bibitem{Brauner1989} M. Brauner, J. S. Briggs, and H. Klar, \href{https://doi.org/10.1088/0953-4075/22/14/010}{J. Phys. B: At. Mol. Opt. Phys. \textbf{22}, 2265 (1989).}
%
\bibitem{Briggs2016} J. S. Briggs, \href{https://doi.org/10.1088/0953-4075/49/17/170501}{J. Phys. B: At. Mol. Opt. Phys. \textbf{49}, 170501 (2016).}
%
\bibitem{Briggs2000} J. S. Briggs and V. Schmidt, \href{https://doi.org/10.1088/0953-4075/33/1/201}{J. Phys. B: At. Mol. Opt. Phys. \textbf{33}, R1 (2000).}
%
\bibitem{Wehlitz1991} R. Wehlitz, F. Heiser, O. Hemmers, B. Langer, A. Menzel, and U. Becker, \href{https://doi.org/10.1103/PhysRevLett.67.3764}{Phys. Rev. Lett. \textbf{67}, 3764 (1991).}
%
\bibitem{Knapp2002b} A. Knapp, M. Walter, Th. Weber, A. L. Landers, S. Sch\"ossler, T. Jahnke, M. Sch\"offler, J. Nickles, S. Kammer, and O. Jagutzki, \href{https://doi.org/10.1088/0953-4075/35/23/103}{J. Phys. B: At. Mol. Opt. Phys. \textbf{35}, L521 (2002).}
%
\end{thebibliography}
\end{document}